\documentclass[usenatbib,usegraphicx]{mn2e}

\newcommand{\OVIdblt}{{O}\kern 0.1em{\sc vi}~$\lambda\lambda 1032, 1038$}
\newcommand{\CII}{\hbox{{C}\kern 0.1em{\sc ii}}}
\newcommand{\CIII}{\hbox{{C}\kern 0.1em{\sc iii}}}
\newcommand{\CIV}{\hbox{{C}\kern 0.1em{\sc iv}}}
\newcommand{\HI}{\hbox{{H}\kern 0.1em{\sc i}}}
\newcommand{\Lya}{\hbox{{Ly}\kern 0.1em$\alpha$}}
\newcommand{\Lyb}{\hbox{{Ly}\kern 0.1em$\beta$}}
\newcommand{\Lyg}{\hbox{{Ly}\kern 0.1em$\gamma$}}
\newcommand{\Lyd}{\hbox{{Ly}\kern 0.1em$\delta$}}
\newcommand{\Lye}{\hbox{{Ly}\kern 0.1em$\epsilon$}}
\newcommand{\Lyz}{\hbox{{Ly}\kern 0.1em$\zeta$}}
\newcommand{\Lyeta}{\hbox{{Ly}\kern 0.1em$\eta$}}
\newcommand{\MgII}{\hbox{{Mg}\kern 0.1em{\sc ii}}}
\newcommand{\OVI}{\hbox{{O}\kern 0.1em{\sc vi}}}
\newcommand{\OVII}{\hbox{{O}\kern 0.1em{\sc vii}}}
\newcommand{\OVIII}{\hbox{{O}\kern 0.1em{\sc viii}}}
\newcommand{\NV}{\hbox{{N}\kern 0.1em{\sc v}}}
\newcommand{\SiII}{\hbox{{Si}\kern 0.1em{\sc ii}}}
\newcommand{\SiIII}{\hbox{{Si}\kern 0.1em{\sc iii}}}
\newcommand{\SiIV}{\hbox{{Si}\kern 0.1em{\sc iv}}}
\newcommand{\FeII}{\hbox{{Fe}\kern 0.1em{\sc ii}}}
\newcommand{\kms}{\ensuremath{\mathrm{km~s^{-1}}}}
\newcommand{\cmsq}{\ensuremath{\mathrm{cm^{-2}}}}
\newcommand{\cm}{\ensuremath{\mathrm{cm}}}
\newcommand{\cc}{\ensuremath{\mathrm{cm^{-3}}}}
\newcommand{\Ang}{\hbox{\textrm{\AA}}}
\newcommand{\mAng}{\hbox{\textrm{m\AA}}}
\newcommand{\Msun}{\ensuremath{M_{\sun}}}
\newcommand{\Zsun}{\ensuremath{Z_{\sun}}}
\newcommand{\NeIX}{\hbox{{Ne}\kern 0.1em{\sc ix}}}
\newcommand{\OI}{\hbox{{O}\kern 0.1em{\sc i}}}
\newcommand{\ergccs}{\ensuremath{\mathrm{erg~cm^{3}~s^{-1}}}}
\newcommand{\kpc}{\ensuremath{\mathrm{kpc}}}
\newcommand{\Kelvin}{\ensuremath{\mathrm{K}}}

\title[C III Absorber Towards PHL 1811]{The $z=0.0777$ \CIII~Absorber Towards PHL 1811 as a Case Study of a Low Redshift Weak Metal Line Absorber}
\author[B. C. Lacki and J. C. Charlton]{B.~C.~Lacki$^1$ and J.~C.~Charlton$^2$\\
$^1$Department of Astronomy, 4055 McPherson Laboratory, 140 West 18th Avenue, Ohio State University, Columbus, OH 43210, USA\\
$^2$Department of Astronomy \& Astrophysics, 525 Davey Laboratory, Pennsylvania State University, University Park, PA 16802, USA}

\begin{document}

\maketitle

\begin{abstract}
We consider the physical conditions and origin of the $z=0.0777$ absorption system observed in \CIII, \CII, \SiIII, \CIV, \OVI, and \HI~absorption along the line of sight towards the quasar PHL 1811.  We analysed the {\it HST}/STIS and FUSE spectra of this quasar and compared the results to Cloudy photoionization and collisional ionization models in order to derive densities, temperatures, and metallicities of the absorbing gas. The absorption can be explained by two \CIII~clouds, offset by 35~{\kms} in velocity, with metallicities of $\sim$one-tenth the solar value. One cloud has a density of order $n_H = 1.2^{+0.9}_{-0.5} \times 10^{-3}$~{\cc} (thickness $0.4^{+0.3}_{-0.2}$~kpc) and produces the observed \CII~and \SiIII~absorption, while the other has a density of order $n_H = 1.2^{+0.9}_{-0.5} \times 10^{-5}$~{\cc} (thickness $80^{+70}_{-40}$~kpc) and gives rise to the observed weak \CIV~absorption.  Cloud temperatures are $\sim 14,000^{+3000}_{-2000}$~K and $\sim 34,000_{-4000}^{+2000}$~K for photoionized models.  Although collisionally ionized clouds with $T \sim 70,000$ K are possible, they are less likely because of the short cooling time-scales involved.  Previous studies revealed no luminous galaxy at the absorber's redshift, so it is probably related to tidal debris, ejected material, a dwarf galaxy, or other halo material in a galaxy group.  Our models also indicate that one of the two clouds would produce detectable weak \MgII~absorption if spectral coverage of that transition existed.  We predict what the system would look like at $z \sim 1$ when the ionizing background radiation was more intense.  We find that at $z \sim 1$ the denser component resembles a \CIV~absorber.  The second \CIII~cloud in this $z=0.0777$ absorber may be analogous to a subset of the more diffuse \OVI~absorbers at higher redshift.
\end{abstract}


\section{Introduction}
\label{sec:Intro}

Quasar absorption lines are an important tracer of the gas in and around galaxies, and in the intergalactic medium (IGM).  Many important diagnostic lines lie in the far-ultraviolet, however, so that studies of low redshift absorbers require such observations. The Space Telescope Imaging Spectrograph (STIS) on the \emph{Hubble Space Telescope} and the \emph{Far Ultraviolet Spectroscopic Explorer} (FUSE) \citep{Moos00} have provided high resolution UV spectra along many lines of sight, covering multiple transitions for many low redshift absorption systems.  Having access to transitions with a range of ionization states allows us to study gas with a wide range of densities ($\sim 10^{-5}$--$100$~{\cc}) in order to probe a large fraction of the volume of the universe.

On the high density end of the scale, the damped {\Lya} absorbers (with $\log N(\HI) > 20.3$~\cmsq) produce strong, saturated absorption in many neutral and singly ionized species.  They are associated with the densest regions in a variety of structures at low redshift, including regions in dwarf and low surface brightness galaxies as well as in luminous galaxies \citep{Rao00,Zwaan05}.  Most of the cross section of luminous galaxies gives rise to Lyman limit ($17.2~\cmsq < \log N(\HI) < 20.3~\cmsq$) and strong \MgII~absorption \citep{Steidel95,Churchill00}. Those \Lya~absorbers with $\log N(\HI) < 17.2~\cmsq$ are collectively called the \Lya~forest \citep[for example, see][]{Charlton00}, and these are responsible for $\sim 30$\% of the baryonic content of the universe \citep*{Penton04} at low redshift, down from $\sim 90$\% at $z \sim 2.5$ (see e.g. \citealt*{Simcoe02}).  Most of the \Lya~forest clouds have $\log N(\HI) < 14.5~\cmsq$ and many are expected to be diffuse (overdensities $\delta$ of order 5) and photoionized \citep{Dave99}.  Other populations of absorbers also exist, associated with galaxy haloes and environments, that are denser and with different sources of ionization \citep{Dave99}, including \OVI, \CIV, and weak \MgII~absorbers.  These are important because they trace regions of emerging structure formation where processes like interactions and feedback shape the properties of intergalactic gas.

Another important species is \OVI, a possible tracer of the shock-heated Warm-Hot Intergalactic Medium (WHIM), which would generally have temperatures of $10^5~K$ to $10^7~K$ \citep{Kang05} and overdensities of $10 \le \delta \le 30$ \citep{Charlton00}. If collisionally ionized, \OVI~lines would trace gas at $10^5~K \le T \le 10^6~K$, particularly near $10^{5.5}~K$. (see e.g. \citealt{Heckman02}, \citealt{Danforth05}).  Simulations predict that the WHIM density would be $\Omega_{WHIM} = 0.3 \Omega_{b}$ \citep{Cen99}.  Other simulations done by \citet{Kang05} suggest the presence of a WHIM component with temperatures of less than $10^5~K$.  This component would be shock-heated and would be arranged into sheets \citep{Kang05}.  Most of it would have $10^4~K < T < 10^5~K$.  The cooler WHIM would comprise $13\%$ of the baryons \citep{Kang05}.  Searching for the WHIM has become a key issue in accounting for local baryons, and \OVI~is an important tracer; however the origin of \OVI~absorbers is likely to be varied, with some likely to be cooler and photoionized, but with some \OVI~absorbers also consistent with higher temperature, collisionally ionized gas and/or with multiphase conditions \citep[Tripp, Savage, \& Jenkins 2000;][]{Heckman02,Berg02,Savage02,Simcoe02,Prochaska04}.  Specifically, \citet{Tripp08} have analysed a sample of 51 $z<0.5$ intervening \OVI~absorbers, with a particular emphasis on a comparison between the \OVI~and the \HI~absorption.  The majority of \OVI~absorbers with aligned \OVI~and \HI~ components tend to be relatively cool ($T<10^5$~K), and consistent with photoionized gas and not with collisional ionization.  However, roughly half of the \OVI~absorbers also have evidence for multiple phases, with the possibility of a hotter, collisional component in addition to the cooler one.  Other diagnostics have been proposed for gauging the WHIM density, such as broad Lyman absorption \citep{Richter06}, and other transitions like \OVII, \OVIII, and \NeIX~(e.g. \citealt{Prochaska04}).

\citet*{Chen01} have found that \CIV~absorbers with $W_r(1548) \ge 0.1~\Ang$ are identified with haloes of bright galaxies, extending to $100 h^{-1}(L/L_B)^{0.5}~\kpc$, with a covering factor of nearly unity.  They proposed that these \CIV~haloes were formed from debris stripped from satellite galaxies.  Roughly half of these \CIV~systems at low redshift also have weak \MgII~detected, which is typically housed in a thinner, higher density layer ($n_e\sim 0.01 - 0.1$~\cc) within the lower density, higher ionization region that gives rise to the \CIV~\citep{Milutinovic05}.  A survey of weak \MgII~absorbers, using \SiII~and \CII~as representative tracers, found $d{\cal N}/dz = 1.00\pm0.20$ for weak \MgII~absorbers with $0.02 < W_r(2796) < 0.3$~{\AA} at $z < 0.4$. Thus these absorbers are only a small fraction of \OVI~absorbers. Weak \MgII~absorbers at $z\sim 1$ have typical metallicities $>0.1$~solar and even greater than solar in many cases, despite their location far from the luminous part of bright galaxies \citep{Charlton03,Narayanan08}.  They have been proposed to arise in extragalactic analogs of Milky Way high velocity clouds, in superwind structures, and generally in metal-rich regions in the cosmic web \citep{Narayanan07,Narayanan08}.

Clearly, there is overlap between the different classes of absorbers, especially in view of phase structure (gas with two different densities at similar velocity) being common.  However, there is a clear progression with \MgII, \CIV, and \OVI~tracing progressively lower density photoionized gas ($n_e\sim 0.1$ -- $10^{-5}~$\cc), but with \OVI~also sometimes arising in collisionally ionized gas at higher temperature.  Thus dense structures give rise to low ionization gas and diffuse structures to highly ionized gas, providing a convenient way to map the distribution of gas in and around galaxies.  An important complication to this picture occurs due to the evolving extragalactic background radiation incident on these systems.  For gas at the same density, the ionization parameter, $\log U = \log n_{\gamma} - \log n_e$, decreases by about an order of magnitude from $z=1$ to $z=0$, leading to less ionized structures in the present than in the past.  For example, \citet{Narayanan05} found that clouds that produced \CIV~absorption and no detected \MgII~($W_r(2796)<0.02$\AA) at $z\sim1$ would produce weak, but detectable \MgII~absorption at $z\sim 0$.  Thus the fundamental nature of the population of absorbers for a given line has changed with time, even if the density distribution of structures does not change.

At low redshift, \CIII~977 provides an important diagnostic of the physical conditions of gas since it is such an intrinsically strong line for gas with an intermediate ionization state, i.e. for $10^{-3.5}~\cc \le n_H \le 10^{-1.0}~\cc$ \citep{Chaffee86}, or an intermediate temperature, i.e. for $T \sim 10^5~K$ \citep{Lehner06}. \citet{Danforth08} has previously considered the number of absorption systems with \CIII~977 absorption lines.  They found that these systems are not as common as those with \OVIdblt~lines with the same equivalent width.  For $d{\cal N}/dz$, the number of systems per unit redshift, \citet{Danforth06} find $d{\cal N}_{\CIII}/dz = 10^{+3}_{-2}$ for $W_{\CIII} \ge 30~m\Ang$, compared to $d{\cal N}_{\OVI}/dz = 17 \pm 3$ for $W_{\OVI} \ge 30~m\Ang$ \citep{Danforth06}.  \citet{Danforth05} argue that the large range of $\log N(\HI)$, and the correlation of $N(\HI)/N(\OVI)$ with $N(\HI)$ imply two phases in many \OVI~absorption systems.  Their model involves a `warm neutral medium' (WNM) core with $10^{3.5}~K < T < 10^{4.5}~K$, which would often produce \CIII~absorption, and a WHIM sheath giving rise to \OVI~absorption \citep{Danforth05}.

The $z=0.0777$ absorber along the line of sight towards the quasar PHL~1811 \citep{Jenkins03} is a \CIII~absorber in which \CII~and \CIV~are only weakly detected, thus we expect it has an intermediate ionization state.  Similar systems include the $z = 0.1736$ system towards PG 1116+215 \citep{Sembach04} and the $z = 0.96463$ system towards PG 1634+706 that we have identified in archival {\it HST}/STIS and Keck/HIRES spectra of that quasar.  In this paper, we aim to constrain the physical conditions of the PHL~1811 \CIII~absorber, using the excellent coverage of many metal line transitions and of the Lyman series lines, with {\it HST}/STIS and FUSE.  This system was among the 51 intervening \OVI~absorbers studied by \citet{Tripp08}, who found three components of absorption with differing ionization states offset by $\sim 30$~{\kms}, but did not consider it specifically in great detail.   In this paper, for the $z=0.0777$ system toward PHL~1811, we perform Cloudy \citep{Ferland00} photoionization modelling and consider whether collisional or photoionization is more likely.  We compare the density, temperature, and metallicity to those of weak \MgII, \CIV, and \OVI~absorbers at low and intermediate redshift in order to see how dominant \CIII~systems fit in.  We then consider information from the \citet{Jenkins03} imaging of the field and discuss which part of the galactic environments and intergalactic medium this system and other systems like it might be probing.

The data reduction and modelling methods are covered in \S~\ref{sec:Methods}.  In section \S~\ref{sec:Spectrum} we present the relevant metal line transitions and Lyman series lines used to constrain the properties of the $z=0.0777$ system towards PHL~1811.  We describe in \S~\ref{sec:2CloudResults} the results of our Cloudy modelling and give the likely physical conditions of the system.  The implications for the origin and environment of the $z=0.0777$ system towards PHL~1811 and other similar systems are discussed in \S~\ref{sec:Discussion}. 

\section{Data and Methods}
\label{sec:Methods}
\subsection{Reduction of the spectra}
\label{sec:Reduction}

We reduced the PHL 1811 spectra from both STIS and FUSE.  PHL~1811 was observed with the E140M grating of STIS ($0.2\arcsec \times 0.06\arcsec$ aperture) for a total of 18.4~ks in program 9418 (PI Jenkins; \citealt{Jenkins03}).  The wavelength coverage was 1150--1730~{\AA}.  Our STIS data reduction is described in \citet{Narayanan05}.  Briefly, the $R=45,000$ E140M data were downloaded, and the pipeline was run on the data \citep{Brown02,Narayanan05}.  Because of the low $S/N$ of individual exposures, the averaging of exposures was a simple weighting by exposure times, as in \citet{Narayanan05}.

\begin{table}
\caption{FUSE observations used.}
\label{table:FUSEObservations}
\begin{tabular}{llccc}
\hline
Data set & Date & Exp & \multicolumn{2}{c}{Counts} \\
 & & s & lif1a & lif1b\\
\hline
P20711010 & 2001 June 22 & 10902 & 53330 & 46507 \\
P10810010 & 2003 June 2 & 15507 & 25410 & 18960 \\
P10810020 & 2003 June 3 & 24255 & 206760 & 149952 \\
P10810030 & 2003 June 3 & 21779 & 173410 & 123660 \\
\hline
\end{tabular}
Data sets are listed in chronological order of observation.
`Exp' is the total exposure time for all exposures in the data set, while `Counts' is the total counts in the channel (lif1a or lif1b) for all exposures in the data set.
\end{table}

Similarly, FUSE data were downloaded from the CalFUSE pipeline.  They had already been processed with CALFUSE version 3.0.  FUSE spectra come in four channels (lif1, lif2, sic1, sic2), which we treated separately because these channels spanned different wavelength bands, including 912--1187~{\AA}.  There were several exposures per dataset, and four datasets for PHL 1811.  The data sets and their properties are listed in Table~\ref{table:FUSEObservations}. We averaged together the exposures for each data set.  The averaging was weighted according to counts.  We then binned each of the resulting spectra by 0.025 \Ang.  Because the actual velocity resolution is of order $v \sim 15~\kms$, the spectral resolution is $R \sim 25,000$ \citep{Sahnow00}.  Thus, the binning did not hamper our analysis.  Then, we were able to average the data sets for each channel.  Once the averaging was completed, we continuum fit each channel of the PHL 1811 FUSE spectrum.

\subsection{The $z=0.0777$ system towards PHL~1811}
\label{sec:Spectrum}

\begin{figure}
\includegraphics[width=80mm]{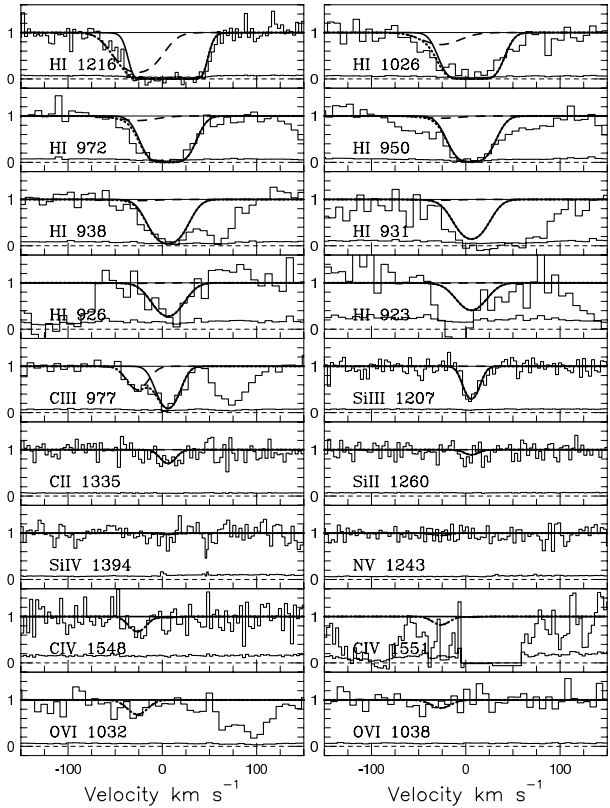}
\caption{Our two cloud model of the $z = 0.0777$ absorber towards PHL 1811, superimposed on the UV spectra.  Transitions shortward of 1150~{\AA} in the rest frame were observed by FUSE, while those longward were observed by STIS.  Both clouds for the superimposed model are photoionized, cloud A having $\log~Z = -1.0$ and $\log~U = -3.0$ (solid model curve), while cloud B has $\log~Z = -1.0$ and $\log~U = -1.0$ (long-dashed model curve).  The combined model, including both clouds, is shown as a dotted curve.}
\label{fig:S4S6}
\label{fig:ABAttr}
\end{figure}

We present the spectrum covering the regions of selected transitions for the $z=0.0777$ absorber towards PHL~1811 in Fig.~\ref{fig:S4S6}, from both the STIS and FUSE spectra.  We aligned the data from the two instruments by finding the average redshifts of the \Lyb, \Lyg, \Lyd, \Lye, and \Lyz~lines in the FUSE spectrum as compared to the \Lya~line in the STIS spectrum, finding that the FUSE spectrum was offset by $\Delta v = 14$~\kms~blueward relative to STIS.  We therefore shifted the FUSE data $14$~\kms~redwards so that transitions covered by the different data sets could be simultaneously modelled.  This shift agrees, within errors, with that applied by \citet{Tripp08}, who compared Galactic \CII~and \FeII~ions observed with two different transitions in the FUSE and STIS spectra.

\begin{table*}
\begin{minipage}{170mm}
\caption{Absorption line properties from Voigt profile fits and models.}
\label{table:VoigtFits}
\begin{tabular}{lcccccccc}
\hline
Lines & \multicolumn{4}{c}{Voigt Profile Fit} & \multicolumn{2}{c}{Model Cloud A} & \multicolumn{2}{c}{Model Cloud B}\\
& $v$ & $\log N$ & $b$ & $W_r$ & $\log N$ & $b$ & $\log N$ & $b$ \\
 & $\kms$ & $\cmsq$ & $\kms$ & \Ang & $\cmsq$ & $\kms$ & $\cmsq$ & $\kms$ \\
\hline
Ly~$\alpha,\beta,\delta,\eta$ & $7 \pm 1$ & $16.00 \pm 0.05$ & $23.0 \pm 0.6$ & $0.47 \pm 0.01$ & 15.8 & 17.7 & 13.8 & 17.3\\
\CIII~977~(\emph{blue}) & $-28 \pm 2$ & $12.87 \pm 0.07$ & $5.6 \pm 4.6$ & $0.030 \pm 0.020$ & ... & ... & 11.6 & 11.6\\
\CIII~977~(\emph{red}) & $6 \pm 1$ & $13.58 \pm 0.03$ & $15.5 \pm 1.5$ & $0.116 \pm 0.007$ & 13.6 & 10.3 & ... & ...\\  
\CII~1335 & $4 \pm 2$  & $13.48 \pm 0.04$ & $27.2 \pm 3.1$ & $0.055 \pm 0.005$ & 13.2 & 10.3 & 11.1 & 11.6\\
\CIV~1548, 1551 & $-25 \pm 2$ & $13.11 \pm 0.09$ & $9.2 \pm 2.6$ & $0.043\pm 0.008$ & 11.7 & 10.3 & 13.0 & 11.6\\
\SiII~1260, 1193, 1190 & $0 \pm 1$  & $12.09 \pm 0.07$ & $5.8 \pm 1.5$ & $0.0152 \pm 0.002$ & 11.8 & 9.7 & 7.4 & 11.2\\
\SiIII~1207 & $6 \pm 1$ & $12.64 \pm 0.03$ & $9.7 \pm 0.7$ & $0.0615 \pm 0.002$ & 12.6 & 9.7 & 8.9 & 11.2\\
\SiIV~1394, 1403 & ... & ... & ... & $<0.0050$ & 11.4 & 9.7 & 9.1 & 11.2\\
\NV~1239, 1243 & ... & ... & ... & $<0.0071$ & 10.1 & 10.1 & 12.6 & 11.5\\
\OI~1302 & ... & ... & ... & $<0.0084$ & 11.3 & 10.0 & 6.9 & 11.4 \\
\OVI~1032 & $-40 \pm 4$ & $13.74 \pm 0.06$ & $32.0 \pm 6.1$ & $0.061 \pm 0.008$ & 8.7 & 10.0 & 13.4 & 11.4\\
\hline
\end{tabular}
\\No components could be fit to \SiIV, \NV, or \OI~so $3\sigma$ equivalent width limits are given.
The rest-frame equivalent widths $W_r$ are given for the strongest listed line for the Voigt profile fit to each species.  \Lya~is heavily saturated, and the Voigt profile column density for \HI~may be inaccurate.  The velocites of the components are with respect to the midpoint of the integrated absorption of the system ($z_{\rm STIS} = 0.077754$).
\end{minipage}
\end{table*}

Table~\ref{table:VoigtFits} gives measured equivalent widths and the results, column densities, Doppler parameters, and central redshifts of Voigt profile fits to detected transitions at $z=0.0777$ (based on an F-test of $\chi^2$ fitting results using the Minfit program; \citealt{Churchill01}).  It also gives equivalent width limits for transitions that fall in a clean part of the spectrum for which limits provide important constraints on our modelling.  All the Voigt profile fits are single component Voigt profile fits, except for \CIII~977 which demands a second component offset by 34~{\kms}.  Some transitions, such as the redward core of the \CIII~977 absorption, are saturated, so that the column densities may be inaccurate.  In particular, the column density in the heavily saturated~\Lya may be underestimated, although~\Lyeta~is not saturated \citep[c.f.][]{Tripp08}.  The \CIII~is the strongest detected metal-line transition in this system.  There is an apparent asymmetry in the \CIII~line, requiring two components with Voigt profile fitting.  Fig.~\ref{fig:ABAttr} shows clearly that the redward component of \CIII~977 is aligned with the observed \CII~1335 and \SiIII~1207 absorption, while the blueward component of \CIII~977 is aligned with the strongest detected \CIV~1548 absorption and the \OVI~1032 absorption.  There also appears to be a small, narrow \CIV~1548 feature at the same redshift as the redward \CIII~977 component, detected with $4.6 \sigma$ significance and rest-frame equivalent width of $0.015 \pm 0.010 \Ang$.

The two component Voigt profile fit to \CIII~977 is also listed in Table~\ref{table:VoigtFits}.  Fig.~\ref{fig:S4S6} shows that the region where \CIV~1551 would be detected is affected by a blend, so that it cannot be considered.  We therefore used models to constrain the properties of two clouds, separated by 34~\kms~in velocity, which we will call Cloud A (redward) and Cloud B (blueward).  \footnote{As we will see, Cloud A corresponds to the main \HI~component found by \citet{Tripp08} and Cloud B corresponds to their \OVI~component in this system.  According to \citet{Tripp08}, these components are separated by $32 \pm 1~\kms$, which is very close to our estimate of a 34~\kms~separation.  Since Cloud A's redshift is determined by \SiIII, measured with STIS, and Cloud B's redshift is determined by the \CIII, measured with FUSE, this demonstrates that the shift we applied to the FUSE data matches the one used by \citet{Tripp08}.  Similarly, the absolute redshift for Cloud B after the shift is 0.077659, which is within 3~\kms~of the redshift \citet{Tripp08} measured for the \OVI~line, 0.07765.}  Cloud A is associated with most of the \SiIII, \CII, \HI, and other low ionization species.  Cloud B is associated with more of the higher ionization absorption, mainly \CIV~and \OVI.

\subsection{Modelling techniques}
\label{sec:Modelling}

To model both photoionization and collisional ionization for the $z = 0.0777$ system in the PHL 1811 line of sight, we used the Cloudy code (version 94, last described by \citealt{Ferland98}).  We ran Cloudy models for Clouds A and B consistent with their derived Voigt profile \CIII~and \SiIII~column densities, thus `optimizing' the models with respect to the observed lines.  For Cloud A, it is better to use the \SiIII~column density as a constraint because the \CIII~977 line is saturated and an accurate column density cannot be derived.  We used a Haardt \& Madau spectrum \citep{Haardt96} as the ionizing spectrum, where an escape fraction of 10 \% was assumed for UV photons from star-forming galaxies \citep{Haardt01}.  For each model that we run, the ionization parameter, $U = n_{\gamma}/n_{H}$ and metallicity are specified.  At $z=0.0777$, the normalization of the Haardt \& Madau spectrum determines the number density $\log n_H = \log n_{\gamma} - \log U$, so that each model corresponds to a particular gas density.   The parameters, $Z$ and $U$ uniquely determine the column densities of all chemical transitions, for that model, so that a synthetic spectrum can be generated.  To create the spectrum, the Doppler parameter $b$ for each line was calculated from $b^2 = 2kT/m + b_{turb}^2$, where $k$ is Boltzmann's constant, $T$ is temperature, $m$ is the atomic mass of the element, and $b_{turb}$ is a contribution from turbulent motion that is constant for all lines from a given cloud.  For photoionized models the equilibrium temperature is from the Cloudy output, while for collisionally ionized models it is an input parameter for the model (higher than photoionization equilibrium gives).  In either case, using this temperature and the observed Doppler parameter for the optimized line, it is possible to calculate $b_{turb}$ and the Doppler parameter for the other lines.  Models with negative $b_{turb}^2$ were ruled out by hand.  Each synthetic model spectrum was then compared with the observed spectrum.

A solar abundance pattern was assumed for all models.  We also assumed a plane parallel geometry for each cloud.  We ran grids of models, varying metallicity $Z$ and ionization parameter $U = n_{\gamma}/n_{H}$ for photoionization models, or temperature $T$ for collisional ionization models.  Grids were run with a spacing of $0.5$ in both $\log Z$ and $\log U$.  At this spacing, models can be distinguished (i.e. adjacent models do not often produce similarly good fits to the data), thus we estimate our accuracy in these parameters to be roughly $\pm 0.25$~dex.  We use the accuracy of 0.25 dex in $U$ to calculate errors in size and density throughout the paper.

\section{Results}
\label{sec:2CloudResults}

\begin{table*}
\begin{minipage}{130mm}
\caption{Cloud properties.}
\label{table:CloudProperties}
\begin{tabular}{lccccccc}
\hline
 & Line fit & $z$ & $\Delta v$ & $\log~N$ & $b$ & $\log~Z$ & $\log~U$\\
 & & & \kms & \cmsq & \kms & & \\
\hline
Cloud A & \SiIII~1207 & 0.077775 & 6 & 12.64 & $9.7\pm 0.7$ & -1.0 & -3.0 \\
Cloud B & \CIII~977 & 0.077659 & -28 & 12.99 & $11.6\pm 2.3$ & -1.0 & -1.0 \\
\hline
\end{tabular}
\\In this table, $z$ is the redshift of the cloud, $\Delta v$ is the offset with respect to the midpoint of the integrated absorption of the system ($z_{\rm STIS} = 0.077754$), $Z$ is the metallicity (with respect to Solar) in our best-fitting photoionization model, $U$ is the ionization parameter in our best-fitting photoionization model.
Other column densities and Doppler parameters for each cloud are model dependent.
\end{minipage}
\end{table*}

We began with Cloud A, for which the \SiIII~column density is listed in Tables~\ref{table:VoigtFits} and \ref{table:CloudProperties}.  Once the parameters of this cloud are constrained, we refit the \CIII~line to determine the column density for Cloud B, dividing out the contribution to that \CIII~from Cloud A.  Our best-fitting photoionization model is briefly described in Table~\ref{table:CloudProperties}.  We show the contributions of each of the two clouds in this model in Fig.~\ref{fig:S4S6}.  Tables~\ref{table:2PhotPhys} and~\ref{table:2ColPhys} list all viable model parameters for each cloud, assuming photoionization and collisional ionization, respectively. 

\begin{table*}
\begin{minipage}{140mm}
\caption{Physical conditions for photoionized models.}
\label{table:2PhotPhys}
\begin{tabular}{rrrccccc}
\hline
Cloud & $\log~Z$ & $\log~U$ & T & $b_{turb}$ & D & $n_H$ & $n_e$ \\
 & & & K & \kms & \kpc & $\cc$ & $\cc$\\
\hline
A & -1.0 & -3.0 & 13700 & 9.3 & 0.42 & 0.0012 & 0.0013 \\
A & -0.5 & -3.0 & 10200 & 9.4 & 0.15 & 0.0012 & 0.0013 \\
A & -1.5 & -2.5 & 19300 & 9.1 & 4.9 & $3.7 \times 10^{-4}$ & $4.1 \times 10^{-4}$ \\
A & -1.0 & -2.5 & 16500 & 9.2 & 1.6 & $3.7 \times 10^{-4}$ & $4.1 \times 10^{-4}$ \\
A & -0.5 & -2.5 & 11700 & 9.3 & 0.52 & $3.7 \times 10^{-4}$ & $4.1 \times 10^{-4}$ \\
A & 0.0 & -2.5 & 6930 & 9.5 & 0.17 & $3.7 \times 10^{-4}$ & $4.1 \times 10^{-4}$ \\
A & 0.5 & -2.0 & 1470 & 9.7 & 0.23 & $1.2 \times 10^{-4}$ & $1.3 \times 10^{-4}$ \\
A & 1.0 & -2.0 & 375 & 9.7 & 0.055 & $1.2 \times 10^{-4}$ & $1.3 \times 10^{-4}$ \\
B & 0.5 & -2.5 & 1230 & 11.5 & 0.0099 & $3.7 \times 10^{-4}$ & $4.1 \times 10^{-4}$ \\
B & 1.0 & -2.5 & 364 & 11.6 & 0.0033 & $3.7 \times 10^{-4}$ & $4.1 \times 10^{-4}$ \\
B & -0.5 & -2.0 & 15200 & 10.7 & 0.33 & $1.2 \times 10^{-4}$ & $1.3 \times 10^{-4}$ \\
B & 0.0 & -2.0 & 9250 & 11.0 & 0.098 & $1.2 \times 10^{-4}$ & $1.3 \times 10^{-4}$ \\
B & 0.5 & -2.0 & 1470 & 11.5 & 0.028 & $1.2 \times 10^{-4}$ & $1.3 \times 10^{-4}$ \\
B & 1.0 & -2.0 & 375 & 11.6 & 0.0093 & $1.2 \times 10^{-4}$ & $1.3 \times 10^{-4}$ \\
B & -1.0 & -1.5 & 27200 & 9.9 & 6.2 & $3.7 \times 10^{-5}$ & $4.3 \times 10^{-5}$ \\
B & -0.5 & -1.5 & 20000 & 10.3 & 2.0 & $3.7 \times 10^{-5}$ & $4.3 \times 10^{-5}$ \\
B & 0.0 & -1.5 & 13400 & 10.8 & 0.59 & $3.7 \times 10^{-5}$ & $4.3 \times 10^{-5}$ \\
B & 0.5 & -1.5 & 7100 & 11.2 & 0.14 & $3.7 \times 10^{-5}$ & $4.3 \times 10^{-5}$ \\
B & 1.0 & -1.5 & 710 & 11.6 & 0.034 & $3.7 \times 10^{-5}$ & $4.3 \times 10^{-5}$ \\
B & -1.0 & -1.0 & 33500 & 9.4 & 83 & $1.2 \times 10^{-5}$ & $1.4 \times 10^{-5}$ \\
B & -0.5 & -1.0 & 25800 & 10.0 & 25 & $1.2 \times 10^{-5}$ & $1.4 \times 10^{-5}$ \\
B & 0.0 & -1.0 & 19700 & 10.4 & 7.8 & $1.2 \times 10^{-5}$ & $1.4 \times 10^{-5}$ \\
B & 0.5 & -1.0 & 15100 & 10.7 & 2.3 & $1.2 \times 10^{-5}$ & $1.4 \times 10^{-5}$ \\
B & 1.0 & -1.0 & 10900 & 10.9 & 0.55 & $1.2 \times 10^{-5}$ & $1.4 \times 10^{-5}$ \\
\hline
\end{tabular}
\\In this table, $b_{turb}$ is the Doppler parameter of the turbulent motion in the cloud, $D$ is the depth/length of the cloud along the line of sight, $n_H$ is the hydrogen number density, and $n_e$ is the electron number density.
\end{minipage}
\end{table*}

\begin{table*}
\begin{minipage}{130mm}
\caption{Physical conditions for collisionally ionized models.}
\label{table:2ColPhys}
\begin{tabular}{rrrrcccc}
\hline
Cloud & $\log Z$ & $\log U$ & $\log T$ & $b_{turb}$ & D & $n_H$ & $n_e$ \\
 & & & K & \kms & \kpc & $\cc$ & $\cc$\\
\hline
A & -1.0 & -5.0 & 4.7 & 8.0 & 0.0039 & 0.12 & 0.13 \\
A & 0.0 & -5.0 & 4.7 & 8.0 & $3.9 \times 10^{-4}$ & 0.12 & 0.13 \\ 
A & -1.0 & -5.0 & 4.8 & 7.5 & 0.0049 & 0.12 & 0.13 \\
A & 0.0 & -5.0 & 4.8 & 7.5 & $4.9 \times 10^{-4}$ & 0.12 & 0.13 \\ 
B & -1.0 & -5.0 & 4.9 & 5.0 & $9.3 \times 10^{-4}$ & 0.12 & 0.14 \\
B & 0.0 & -5.0 & 4.9 & 5.0 & $9.3 \times 10^{-5}$ & 0.12 & 0.14 \\ 
B & -1.0 & -5.0 & 5.0 & $\la 3.5$ & 0.0016 & 0.12 & 0.14 \\
B & 0.0 & -5.0 & 5.0 & $\la 3.5$ & $1.6 \times 10^{-4}$ & 0.12 & 0.14 \\
\hline
\end{tabular}
\\In this table, $b_{turb}$ is the Doppler parameter of the turbulent motion in the cloud, $D$ is the depth/length of the cloud along the line of sight, $n_H$ is the hydrogen number density, and $n_e$ is the electron number density.  The calculations for the depth and densities assume that $\log U = -5$; other ionization parameters with different densities are also consistent with the constraints, and give different densities and depths.
\end{minipage}
\end{table*}

\subsection{Cloud A}
\label{sec:AResults}
Our Voigt profile fit to \SiIII~sets the redshift of cloud A as $z_{STIS} = 0.077775$.  The column density of \SiIII~is $\log N(\SiIII) = 12.64 \pm 0.03~\cmsq$ and the Doppler parameter is $b(\SiIII) = 9.7 \pm 0.7~\kms$.

\subsubsection{Photoionization}
\label{sec:APhoto}

\begin{figure}
\includegraphics[width=84mm]{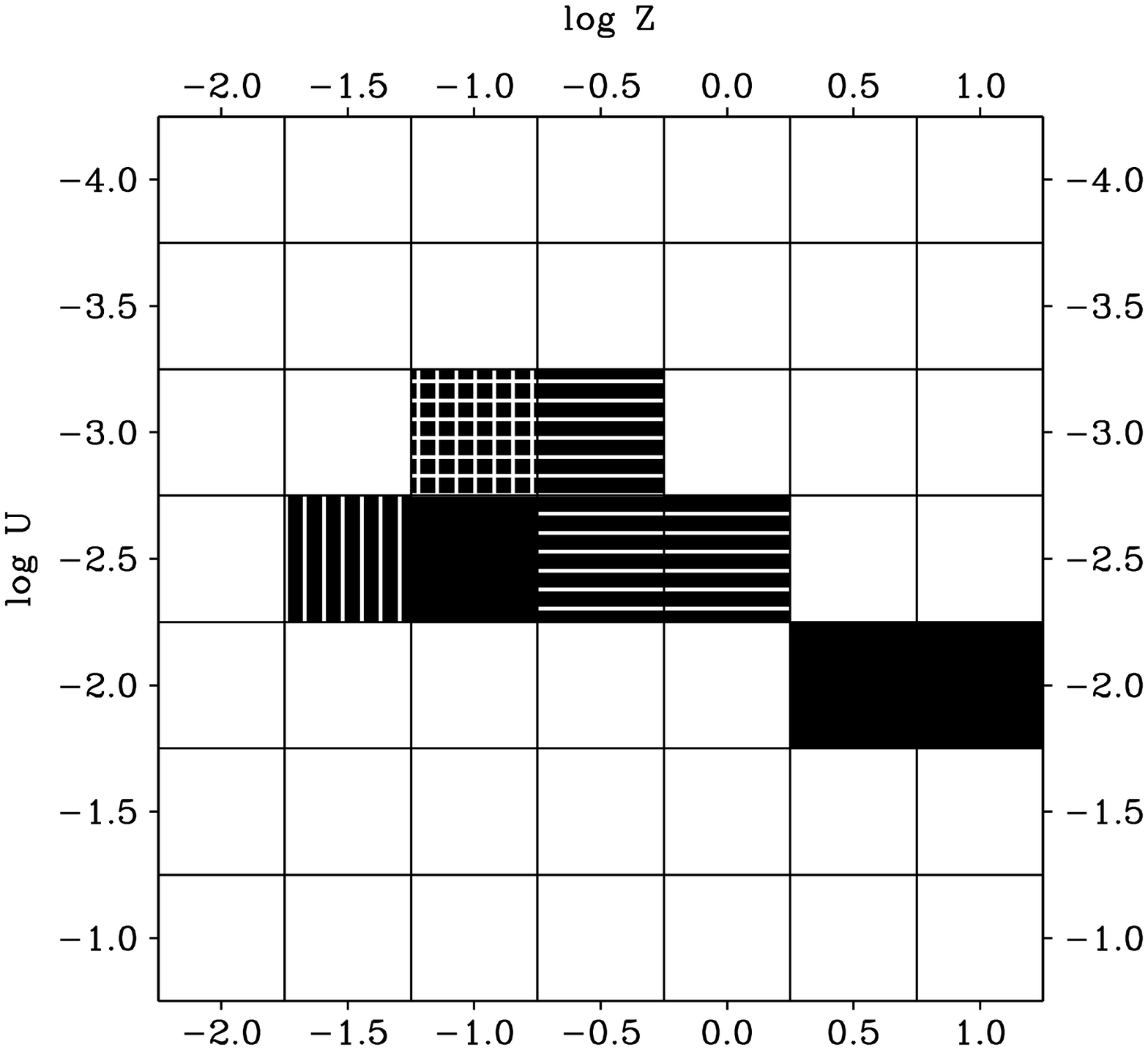}
\caption{This plot represents the grid of photoionization models we ran for cloud A.  Shades boxes are those models that produced the \SiIII~absorption without overproducing other species.  Boxes with horizontal cross-hatching represent models with \SiIII~and \CII; the vertically cross-hatched boxes are models with \SiIII~and \Lya.  Only our one favoured model produced enough \SiIII, \CII, and \Lya.}
\label{fig:APhotGrid}
\end{figure}

We ran a grid of photoionization models with $-4.0 \le \log U \le -1.0$ and $-2.0 \le \log Z \le 1.0$ at 0.5 dex intervals.  We find that for some values of ionization parameter and metallicity, the \SiIII~can be explained without overproducing other species.  Our models indicate that $-3.0 \le \log U \le -2.0$ and $-1.5 \le \log Z$, as seen in Fig.~\ref{fig:APhotGrid}.  At lower ionization parameters ($-3.0 < \log U$), \CII~is always overproduced, often along with \HI~and \SiII.  Higher ionization parameter models overproduce several species, including \CIII, \CIV, \SiIV, \NV, and \OVI.

Cloud A also must fit the redwards wing of the \HI~absorption and the detected \CII~absorption.  Cloud A can explain the red side of the \HI~absorption for two models: when $\log Z = -1.0$ and $\log U = -3.0$, as well as for $\log Z = -1.5$ and $\log U = -2.5$.  The blueward wing of both the \HI~and \CIII~absorption cannot be fit with cloud A alone. Of the two models that match the red side of the \HI~absorption, the \CII~absorption is matched when $\log~U = -3.0$.  To fit both the \HI~and \CII~absorption requires that $\log Z \sim -1.0$ and $\log U \sim -3.0$, so we favour this model among those in our grid.  This model does not predict the very narrow \CIV~feature detected at the same redshift, which may mean that at least one other component is at the same redshift.

The physical conditions predicted by the photoionization models we have considered for cloud A are listed in Table~\ref{table:2PhotPhys}.  Photoionization models predict cloud sizes between $0.055^{+0.043}_{-0.024}$ \kpc~and $4.9^{+3.8}_{-2.1}$ \kpc.  Hydrogen number densities of $1.2^{+0.9}_{-0.5} \times 10^{-4}~\cc$ ($\delta = 470^{+370}_{-210}$) to $0.0012^{+0.0009}_{-0.0005}~\cc$ ($\delta = 4700^{+3700}_{-2100}$) are expected.  Temperatures range from $380~\Kelvin$ to $19000~\Kelvin$. If we require that cloud A fit the red edge of the Lyman series and the \CII~absorption, as in our favoured model, then the density, sizes, and temperatures are constrained to be towards the higher ends of these ranges.  This ($\log U = -3.0$, $\log Z = -1.0$) model indicates a size of $0.42^{+0.33}_{-0.18}$ kpc, a density of $0.0012^{+0.0009}_{-0.0005}$ \cc, and a temperature of $13700^{+2800}_{-1800}~\Kelvin$.

In our favoured model, we predict that the column density of the \HI~from Cloud A is $N(\HI) = 15.77^{+0.31}_{-0.24}~\cmsq$, with a Doppler parameter of $b_H \approx 17.7~\kms$.  To compare, \citet{Tripp08} identified two \HI~lines in the $z = 0.0777$ system towards PHL 1811.  The first was $24 \pm 5~\kms$~redwards of the \OVI~line they identified with $\log N(\HI) = 13.56 \pm 0.09~\cmsq$~and was extremely wide, with $b_H = 71 \pm 12~\kms$ \citep{Tripp08}.  Our derived hydrogen column density and Doppler parameter of Cloud A are inconsistent with this feature.  The second hydrogen absorption component identified by \citet{Tripp08} was $32 \pm 1~\kms$~redwards of the \OVI~line with $\log N(\HI) = 16.03 \pm 0.07~\cmsq$ and $b_H = 19 \pm 1~\kms$.  The Doppler parameter of this feature is in good agreement with our predicted Doppler parameter for our favoured model for Cloud A, and we identify this \HI~component with Cloud A.  Our predicted \HI~column density is somewhat less than measured by \citet{Tripp08}, which may mean that there are other clouds near this redshift causing additional \HI~absorption, as also suggested by the \CIV~feature at this redshift.  However, the difference is not very great considering the errors.

\subsubsection{Collisional ionization}
\label{sec:AColIon}

\begin{figure}
\includegraphics[width=84mm]{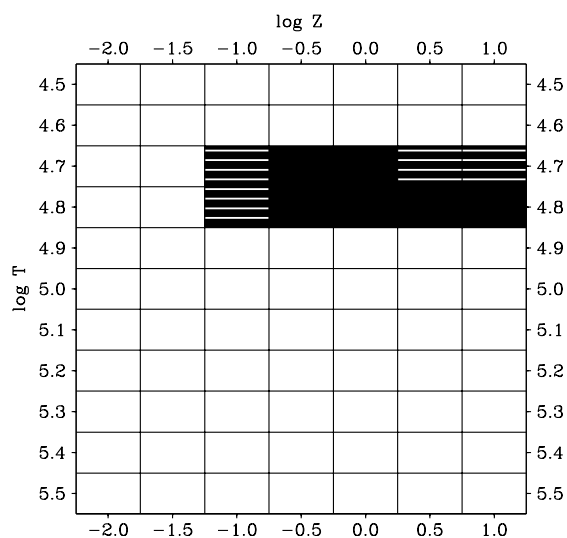}
\caption{This plot represents the grid of collisional ionization models we ran for cloud A.  Shaded boxes are those models that produced the \SiIII~absorption without overproducing other species.  Horizontally cross-hatched boxes represent models with \SiIII~and \CII.  No allowed model produced enough \Lya.}
\label{fig:ACollGrid}
\end{figure}

We also tried collisional ionization for temperature in the ranges of $1.0 \le \log T \le 5.5$.  The temperature was set to be constant in Cloudy.  We also assumed $\log U = -5.0$, so that the ionizing background was insignificant and the models were effectively in collisional ionization equilibrium (CIE).  However, any sufficiently small ionization parameter would produce similar results, leading to a degeneracy in these models in density and size.  Our grid used one dex intervals for $\log T \le 4.5$ and 0.1 dex intervals for $4.5 \le \log T \le 5.5$.  We used metallicities ranging from $\log Z = -2$ to $\log Z = 1$, in 0.5 dex intervals.  We found that the collisional ionization could explain the absorption for $4.7 \le \log T \le 4.8$, if $\log Z \ge -1.0$ (see also Fig.~\ref{fig:ACollGrid}).  Lower temperatures overproduce \CII~and sometimes other lower ionization species, while higher temperatures overproduce \CIII, \CIV, \SiIV, and sometimes other higher ionization species.  Models with $\log~T = 4.7$ matched the \CII~absorption.  For all models consistent with constraints from the metal line transitions, the \HI~absorption was underproduced.  However, \Lya~was overproduced when $\log Z \le -1.5$, so it may be the case that \Lya~can be fit if $-1.5 \le \log Z \le -1.0$.  Models with $\log Z = -1.5$ still do not produce enough \Lyg~and \Lyd~absorption, though, in contrast to photoionization models (which do reproduce the \Lyg~and \Lyd, as demonstrated in Fig.~\ref{fig:S4S6}).  An even lower metallicity would be needed to account for the \Lyg~and \Lyd, which is already ruled out because of the \Lya~absorption.  Since collisional ionization cannot reproduce these lines, we favour photoionization.

The physical conditions predicted by our collisional ionization models for cloud A are listed in Table~\ref{table:2ColPhys}.  The hydrogen density for collisional ionization models is $n_H \ga 0.0012~\cc$, assuming that $\log U \la -3$ so that photoionization becomes unimportant, since the number density of extragalactic background photons is constant at that redshift.  Sizes are given by $4.8 \times 10^{-5} (n_H / \cm^{-3})^{-1}$ to $4.8 \times 10^{-4} (n_H / \cm^{-3})^{-1}$ \kpc. 

\subsection{Cloud B}
\label{sec:BResults}
Cloud B was fit to the remaining bluewards \CIII~977 absorption, after dividing out the absorption predicted for cloud A with the $\log Z = -1.0$, $\log U = -3.0$ model.  Our Voigt profile fit indicates that cloud B is located at $z_{STIS} = 0.077659$.  The column density of \CIII~is found to be $\log N(\CIII) = 12.99~\cmsq$, and our measured Doppler parameter is $b(\CIII) = 11.6 \pm 2.3~\kms$.

\subsubsection{Photoionization}
\label{sec:BPhoto}

\begin{figure}
\includegraphics[width=84mm]{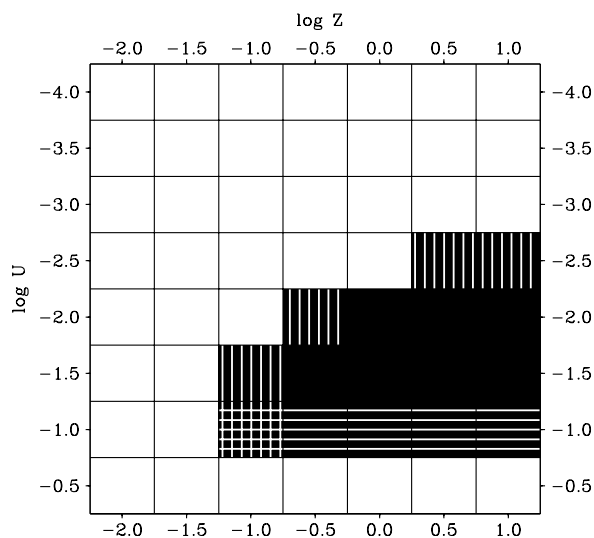}
\caption{This plot represents the grid of photoionization models we ran for cloud B.  Shaded are those models that produced the \CIII~absorption without overproducing other species.  Horizontally cross-hatched boxes represent models with \CIII~and \CIV; vertically cross-hatched boxes are models with \CIII~and \Lya.  Only one model, our favoured model, produced enough \CIII, \CIV, and \Lya.}
\label{fig:BPhotGrid}
\end{figure}

For cloud B, we used the same grid as cloud A, but expanded to include $\log U = -0.5$.  We found that cloud B did not overproduce metal line or Lyman series transitions when $-2.5 \le \log U \le -1.0$ and $-1.0 \le \log Z$.  Lower ionization parameter models had too much \SiIII~or lower ionization species.  Fig.~\ref{fig:BPhotGrid} shows exactly which models were consistent with the spectrum.  

However, if we require that Cloud B explains two additional features, the bluewards \HI~wing and the aligned \CIV~1548 absorption, a specific model is favoured.  To match the ratio of \CIV~1548 to \CIII~977 the ionization parameter is constrained to be $\log U = -1.0$.  For this value of $\log U$, a metallicity of $\log Z = -1.0$ matches the red \HI~wing.  Thus our favoured model for Cloud B, among the choices on the grid, has $\log U = -1.0$ and $\log Z = -1.0$.

The physical conditions for considered photoionization models of cloud B are listed in Table \ref{table:2PhotPhys}.  Photoionization models predict sizes ranging from $0.0033^{+0.0026}_{-0.0014}$ \kpc~to $83^{+65}_{-36}$ \kpc, densities from $1.2^{+0.9}_{-0.5} \times 10^{-5}~\cc$ ($\delta = 48^{+37}_{-21}$) to $3.7^{+2.9}_{-1.6} \times 10^{-4}~\cc$ ($\delta = 1500^{+1200}_{-660}$), and temperatures from $360~\Kelvin$ to $33400~\Kelvin$.  However, requiring that cloud B explain both the \HI~and the small \CIV~absorption feature observed leaves only one model on our grid, one which has a high ionization parameter parameter.  This model ($\log U = -1.0$, $\log Z = -1.0$) corresponds to a hydrogen number density of $1.2^{+0.9}_{-0.5} \times 10^{-5}~\cc$.  The size of the model cloud would be $83^{+65}_{-36}$ \kpc, and its temperature would be $T = 33500^{+2300}_{-3900}~\Kelvin$.  If both cloud A and cloud B correspond to our favoured photoionization models, then cloud B has a lower density than cloud A by about two orders of magnitude, and cloud B is much larger than cloud A.

We note that the column density of \OVI~from Cloud B in our favoured model is $\log N(\OVI) = 13.43_{-0.63}^{+0.43}~\cmsq$, with an oxygen Doppler parameter of $b_O \approx 11.4~\kms$.  \citet{Tripp08} also identified an \OVI~line at the same redshift as our Cloud B, with a column density of $\log N(\OVI) = 13.56 \pm 0.10~\cmsq$~and a Doppler parameter of $b_O = 26^{+12}_{-8}~\kms$.  The predicted column density for Cloud B in our favoured model is roughly in agreement with that measured by \citet{Tripp08}.  We therefore identify the \OVI~component measured by \citet{Tripp08} as Cloud B; however, our predicted Doppler parameter is significantly smaller than that measured by \citet{Tripp08}.

\subsubsection{Collisional ionization}
\label{sec:BColIon}

\begin{figure}
\includegraphics[width=84mm]{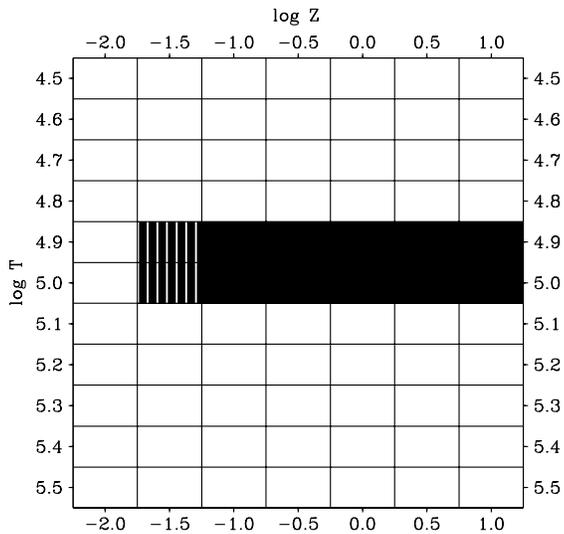}
\caption{This plot represents the grid of collisional ionization models we ran for cloud B.  Shaded boxes are those models that produced the \CIII~absorption without overproducing other absorption.  The vertically cross-hatched boxes represent the model with \CIII~and \Lya.  All allowed models underproduced \CIV.}
\label{fig:BCollGrid}
\end{figure}

As with Cloud A (\S~\ref{sec:AColIon}), we approximated collisional ionization equilibrium by assuming an ionization parameter of $\log U = -5.0$ and some constant temperature $T$.  Our grid of collisional ionization models spanned $1.0 \le \log T \le 5.5$, at 1 dex intervals for $\log T \le 4.0$ and 0.1 dex intervals for $4.5 \le \log T \le 5.5$.  We used metallicities of $\log Z = -2.0$ to $\log Z = 1.0$ in 0.5 dex intervals.  We find that for $\log Z \ge -1.0$, the \CIII~absorption can be explained without overproducing other species if $4.9 \le \log T \le 5.0$ (see also \ref{fig:BCollGrid}).  Lower temperatures overproduce \SiIII~and sometimes other low ionization species.  Higher temperatures overproduce \CIV~and sometimes other high ionization species.  The model with $\log Z = -1.0$ and $\log T = 4.9$ produces the bluewards wing of \Lya.  However, no model produces the observed \CIV~feature.  Because of this, photoionization is favoured for Cloud B.  On the other hand, the \CIV~abundance did increase with temperature, so that some \CIV~did appear at $\log T = 5.0$.  It is possible that there is a solution with $5.0 \le \log T \le 5.1$ where collisional ionization can reproduce all of the characteristics of Cloud B.

The physical conditions predicted by the collisional ionization models we considered for cloud B are listed in Table~\ref{table:2ColPhys}.  All of the collisional ionization models imply that $n_H \ga 1.2 \times 10^{-5}~\cc$, if $\log U \la -1$ so that the photoionizing background is unimportant.  The size is $1.1 \times 10^{-5} (n_H / \cm^{-3})^{-1}$ to $1.9 \times 10^{-4} (n_H / \cm^{-3})^{-1}$ \kpc.

\section{The Context of the $z=0.0777$ Absorber Toward PHL 1811}
\label{sec:Discussion}

\subsection{The IGM phase of the $z=0.0777$ absorber and its ionization}
\label{sec:DiscussionColIon}
Cosmological simulations predict a multiphase structure of the IGM: a very low density, photoionized Lyman alpha forest on the largest scales; a somewhat higher density, shock heated warm-hot ionized medium (WHIM) on smaller scales; and high density, photoionized clouds around galaxies \citep{Dave99,Kang05}.  The WHIM is considered to have temperatures of $T = 10^5~\Kelvin$ to $T = 10^7~\Kelvin$, and so has been sought with high ionization lines like \OVI~(for example, \citealt{Heckman02}).  However, \citet{Kang05} predict a cool WHIM with temperatures above $10^4~\Kelvin$ and densities of order $\sim 10^{-6}~\cc$.  It is also possible that \CIII~could serve as a probe of the interface between phases in multiphase models of \OVI~absorbers \citep{Danforth05}.  It can even be the case that the absorber studied here does not easily fall into any of these categories, and is more directly associated with a galaxy or its immediate surroundings.  Knowing whether the $z = 0.0777$ system towards PHL 1811 is collisionally ionized or photoionized is therefore important to finding its place in the IGM.

In their study of \CIII~absorption systems, \citet{Danforth06} argued that they were photoionized.  First, they found that $N(\CIII)$ is correlated with $N(\HI)$ in \CIII~absorber systems, which suggests a coupling between the two.  However, they note that there is much scatter in the relationship.  Second, they found that the column density distributions of \CIII~absorbers and \Lya~absorbers were similar.  Both could be characterised as power laws proportional to $N^{-\beta}$, where $N$ is the column density with $\beta_{\HI} = 1.68 \pm 0.11$ and $\beta_{\CIII} = 1.68 \pm 0.04$.  Finally, they calculated the cooling times of shock heated \CIII~absorbers and found that they were extremely short.

The statistical arguments for the ensemble of \CIII~systems studied by \citet{Danforth06} are not very useful for the specific system studied here, since we do not know where it lies on the distribution.  However, we may calculate the cooling times of both cloud A and cloud B from our collisional ionization models, using the same approach as \citet{Danforth06}.  In collisional ionization equilibrium, \begin{equation}\tau_{cool} = {3kT\over{n_H\Lambda_N}},\end{equation} where $n_H$ is the hydrogen number density, and $\Lambda_N$ is the cooling rate coefficient \citep{Danforth06,Sutherland93}.  By using the cooling rate coefficients $\Lambda_N$ given in Sutherland \& Dopita \citep{Sutherland93}, we derive the cooling times, which are listed in Table~\ref{table:CoolTimes}~for Clouds A and B.

\begin{table*}
\begin{minipage}{140mm}
\caption{Cooling times for collisionally ionized models.}
\label{table:CoolTimes}
\begin{tabular}{lllrccc}
\hline
$\log U$ & $\log Z$ & $\log T$ & $n_H$ & $\Lambda_N$ & $\tau_{cool}$ & $\tau_{cool}$ \\
 & & \Kelvin & \cc & \ergccs & s & Gyr\\
\hline
0.73 & 0.0 & 4.7 & $2.5 \times 10^{-7}$ & $4.79 \times 10^{-22}$ & $1.8 \times 10^{17}$ & $5.5$ \\
0.73 & -0.5 & 4.7 & $2.5 \times 10^{-7}$ & $2.19 \times 10^{-22}$ & $3.8 \times 10^{17}$ & $12$ \\
0.73 & -1.0 & 4.7 & $2.5 \times 10^{-7}$ & $1.12 \times 10^{-22}$ & $7.4 \times 10^{17}$ & $24$ \\
$\la$-3.0 & 0.0 & 4.7 & $\ga 0.0012$ & $4.79 \times 10^{-22}$ & $4.3 \times 10^{10} \times (n_H / \cm^{-3})^{-1}$ & $1.3 \times 10^{-6} \times (n_H / \cm^{-3})^{-1}$ \\
$\la$-3.0 & -0.5 & 4.7 & $\ga 0.0012$ & $2.19 \times 10^{-22}$ & $9.5 \times 10^{10} \times (n_H / \cm^{-3})^{-1}$ & $3.0 \times 10^{-6} \times (n_H / \cm^{-3})^{-1}$ \\
$\la$-3.0 & -1.0 & 4.7 & $\ga 0.0012$ & $1.12 \times 10^{-22}$ & $1.9 \times 10^{10} \times (n_H / \cm^{-3})^{-1}$ & $5.9 \times 10^{-6} \times (n_H / \cm^{-3})^{-1}$ \\
0.73 & 0.0 & 4.9 & $2.5 \times 10^{-7}$ & $9.77 \times 10^{-22}$ & $1.3 \times 10^{17}$ & $4.3$ \\
0.73 & -0.5 & 4.9 & $2.5 \times 10^{-7}$ & $4.47 \times 10^{-22}$ & $3.0 \times 10^{17}$ & $9.3$ \\
0.73 & -1.0 & 4.9 & $2.5 \times 10^{-7}$ & $2.29 \times 10^{-22}$ & $5.7 \times 10^{17}$ & $18$ \\
$\la$-1.0 & 0.0 & 4.9 & $\ga 1.2 \times 10^{-5}$ & $9.77 \times 10^{-22}$ & $3.4 \times 10^{10} \times (n_H / \cm^{-3})^{-1}$ & $1.1 \times 10^{-6} \times (n_H / \cm^{-3})^{-1}$ \\
$\la$-1.0 & -0.5 & 4.9 & $\ga 1.2 \times 10^{-5}$ & $4.47 \times 10^{-22}$ & $7.4 \times 10^{10} \times (n_H / \cm^{-3})^{-1}$ & $2.3 \times 10^{-6} \times (n_H / \cm^{-3})^{-1}$ \\
$\la$-1.0 & -1.0 & 4.9 & $\ga 1.2 \times 10^{-5}$ & $2.29 \times 10^{-22}$ & $1.4 \times 10^{11} \times (n_H / \cm^{-3})^{-1}$ & $4.6 \times 10^{-6} \times (n_H / \cm^{-3})^{-1}$ \\
\hline
\end{tabular}
For comparison, the mean baryonic number density of the Universe is $2.5 \times 10^{-7}~\cc$.
These cooling times assume that the ionization is from collisional ionization only; at high $U$ (low density), photoionization can keep the cloud in thermal equilibrium.
\end{minipage}
\end{table*}

We find that $\tau_{A,cool} \sim 5000(n_H / \cc)^{-1}~\textrm{yr}$ and $\tau_{B,cool} \sim 2000(n_H / \cc)^{-1}~\textrm{yr}$, for $\log Z \sim -1$.  Collisionally ionized clouds are dense, so they will necessarily have very short cooling times: a few million years or less.  The constraints are more severe for cloud B than cloud A, but they are the same order of magnitude for both.  Even at $\log U = -2$, approaching the photoionization regime, $\tau_{B,cool} \sim 20~\textrm{Myr}$, vastly shorter than time-scales of large scale structure formation.  Only when the density is comparable to the critical density of the Universe ($n_H \sim 2.5 \times 10^{-7} \cc$) are the cooling times comparable to the Hubble time.

Another reason that we prefer a photoionization explanation of the system is the detected \HI~absorption in Cloud A~and \CIV~absorption in Cloud B.  Photoionization models could explain the presence of these species, but our collisionally ionized models could not account for all the observed absorption without overproducing \CIII.  However, it is possible that the system contains otherwise hidden phases that would produce this absorption.  Furthermore, as we discussed in \S~\ref{sec:AColIon} and \S~\ref{sec:BColIon}, it is possible that our grids were too coarse to find collisional ionization solutions.  While we argued against this possibility for Cloud A, because of the \Lyg~and \Lyd~absorption, it remains a possibility for the \CIV~absorption of Cloud B.

Based on these arguments, we conclude that photoionization works better as the source of ionization for these clouds than collisional ionization, especially for Cloud A. The system is not likely to be part of the WHIM, whether cool or warm. That still leaves the question of which photoionized phase of the IGM the absorber is a part of.  Is each component part of the low density \Lya~forest, a high density cloud associated with a galaxy, or even a transitional state with a density normally associated with the WHIM?  From the densities given in Table~\ref{table:2PhotPhys}, it is clear that both clouds are relatively high density.  With a number density of $\sim 10^{-3}~\cc$ (overdensity $\delta \sim 5000$) in our preferred model, cloud A is far too dense to be a part of the \Lya~forest or any photoionized WHIM.  Therefore, it is probably `condensed' and associated with some galaxy \citep{Dave99}.

Cloud B is still far too dense to be a part of the \Lya~forest, with $n_H \sim 10^{-5}~\cc$ ($\delta \sim 50$) in our favoured model, it is fairly dense for the WHIM, and an order of magnitude denser than typical cool WHIM gas.  However, cloud B is also underdense for a typical condensed \Lya~absorber ($\delta > 170$) in the models of \citet{Dave99}.  Because it is probably photoionized and relatively cold ($T \sim 30000~K$), it does not cleanly fall into either the WHIM phase or the condensed phase.  If it is primordial, it may represent some kind of transitional absorber.  This may also be seen from its larger size (about $80^{+60}_{-40}~\kpc$). If cloud B is collisionally ionized, it will be denser, and more clearly associated with an individual galaxy.

This discussion assumes that the absorbers are primordial, in that the gas has not been processed by a galaxy and returned to the IGM, but is still undergoing structure formation.  However, it is also possible that low density gas has been ejected out of a galaxy, perhaps as a wind or as tidal debris.  Then, Cloud B may still be associated with a galaxy, even with its low density, and however it is ionized.  Since Cloud B cannot be clearly identified with the usual IGM phases, a galactic origin of this kind may be preferred.  The possibility of an origin from within a galaxy will be considered more in \S~\ref{sec:Association}.

\subsection{The big picture: association with a galaxy?}
\label{sec:Association}

When they identified the $z = 0.0777$ absorber along the PHL 1811 line of sight, \citet{Jenkins03} proposed that it was a tidal debris cloud.  They noted the existence of absorption systems on this sight line at $z = 0.0734$, $0.0777$, $0.0790$, and $0.0809$ \citep{Jenkins03}.  These are fairly close in redshift to the nearby galaxy G 158, at $z = 0.0808$.  That galaxy lies at a projected distance of $34~h_{70}^{-1}~\kpc$ from the sightline \citep{Jenkins03}.  Another galaxy, G 169, is at $z = 0.0804$ with impact parameter $87~h_{70}^{-1}~\kpc$ \citep{Jenkins03}.  A follow-up ACS image revealed a third bright galaxy coincident with the quasar, probably the quasar host, though it is remotely possible that it lies at another redshift \citep{Jenkins05}.  Because of the relatively few metal lines visible in these systems, they initially proposed that all the clouds are linked with the galaxy G158 and are the result of tidal debris \citep{Jenkins03}.  Galaxy interactions are known to produce large tidal tails of gas and stars, stripped off from the interacting galaxies.  Neutral hydrogen is among the constituents of these tails.  Famous among tidal tails is the Magellanic Stream, which has a velocity gradient of up to $390~\kms$ \citep{Putnam03}.  Other examples include NGC 4676a and b, with a \HI~velocity amplitude of $420~\kms$; and Arp 295 a and b, with a \HI~velocity amplitude reaching $520~\kms$ \citep{Hibbard96}.  Column densities of \HI~can be up to $10^{19}~\cmsq$ in tidal tails, with total \HI~masses of $10^9 \Msun$.  Spatially, they may extend 100 \kpc~\citep*{Bekki05}. On the other hand, the velocity difference for the $z = 0.0777$ system is still rather large, with $\Delta v = -870~\kms$ to G 158 and $\Delta v = -730~\kms$ to G 169 \citep{Jenkins03}.  These large velocities could be problematic for the tidal debris scenario, as stated by \citet{Jenkins03}. 

Another, less dramatic scenario is that the $z = 0.0777$ system towards PHL 1811 is associated with the local large scale structure.  There are a number of galaxies at nearly the same redshift.  One of the nearest is SDSS J215513.70-091805.8 at $z = 0.0745$, which is 5.3' away from PHL 1811 \citep{Merchan05}, implying an impact parameter of $470~h_{70}^{-1} \kpc$.  Several other galaxies cluster at this redshift, including several at $z = 0.078$.   \citet*{Bregman04} investigated the large scale structure at $z \sim 0.08$ on the PHL 1811 sightline.  The supercluster Aquarius B is in its direction at the same redshift \citep{Bregman04}.  \citet{Bregman04} list five clusters in Aquarius B with $0.08 \le z \le 0.09$.  They note that the PHL 1811 sightline passes between two pairs of these clusters: A2376 and A2377, as well as A2402 and A2410 \citep{Bregman04}.  However, the PHL 1811 sightline does not pass near any of the clusters themselves, coming only within 50' \citep{Bregman04}.  Therefore, absorbers on the PHL 1811 sightline could be in cosmic filaments connecting the clusters, but not the clusters themselves.  \citet{Bregman04} note that the velocity dispersions of the $0.07 \le z \le 0.08$ absorbers, up to 2250 \kms~for the $z = 0.07344$ absorber, could arise from motions in superclusters.  SDSS data does not cover PHL 1811, but does extend to near regions in the sky.  From that SDSS data, \citet{Jenkins05} found that there is a filament of galaxies near PHL 1811's sightline that includes two S0 galaxies at $z = 0.080$.

Several studies have found that while strong \Lya~absorbers \citep*{Penton02} and \OVI~absorbers \citep{Stocke06} may be associated with galaxies, weak \Lya~absorbers do not cluster as much.  According to \citet{Lanzetta95}, a third to two-thirds of all strong \Lya~absorbers are within galaxies.  They also found an inverse relationship between the impact parameter and the equivalent width of the \Lya~line \citep{Lanzetta95}.  However, their study focused on strong \HI~absorbers.  Furthermore, they did not find a relationship between equivalent width and type of galaxy, which is what would be expected if galaxies were responsible \citep{Lanzetta95}.  In fact, \citet{Penton02} found that the majority of weak ($W_{r}(1216) < 68~m\Ang$, $N < 10^{13.2}~\cmsq$) absorbers have no bright galaxy within $1~h_{70}^{-1}~\textrm{Mpc}$.  Instead, they appear to fall along filaments connecting the galaxies \citep{Penton02}.  Clouds like these are reported in cosmological simulations \citep{Dave99}.

Perhaps the system at $z = 0.0777$ towards PHL 1811 is part of the large scale structure linking the Aquarius B supercluster, including groups at $z = 0.078$ and the one containing G 158.  The projected distance between SDSS J215513.70-091805.8 and the PHL 1811 line of sight is nearly the median projected distance between galaxies and \Lya~absorbers, $500~h_{70}^{-1}~\kpc$ ($W > 16~m\Ang)$ \citep{Penton02}.  If the PHL 1811 $z = 0.0777$ system is part of a supercluster filament, its redshift near those of SDSS J215513.70-091805.8 and G 158 may be explained.  We should note, however, that the system at $z = 0.0777$ towards PHL 1811 is a relatively strong \Lya~absorber.  In fact, the implied overdensity would be far greater than that typical of a large scale structure associated \Lya~absorber ($\delta \la 10$).  Rather, the temperature and density are more like that of the `condensed' galactic \Lya~clouds, especially for cloud A (see e.g. \citealt{Dave99}).  Cloud B, with a lower overdensity, may indeed be associated with structure on the hundred kiloparsec scale in some way.

A final possibility to consider is that the system at $z = 0.0777$ on the PHL 1811 line of sight is associated with a dwarf galaxy too faint to see.  The dwarf, of course, may be located within a filament of galaxies in the large scale structure.  This could reconcile the high density of PHL 1811 $z = 0.0777$ with the lack of an observed galaxy.  Such a dwarf might even be in the same group as G 158 or G 169.  \citet{Salpeter93} proposed that `vanishing' dwarf galaxies are responsible for \Lya~absorption.  The dwarfs would form late in cosmic history.  An initial burst of star formation will result in supernovae, creating winds that clear out the galaxy centres.  This terminates star formation.  As a result, the galaxy glows only with residual starlight.  However, an outer halo of neutral gas, enriched with metals, remains \citep{Salpeter93}. 

At least one absorber, with similar properties to the $z = 0.0777$ system towards PHL 1811, is likely to be associated with a dwarf galaxy.  \citet{Stocke04} found an impact parameter of $70~h_{70}^{-1}~\kpc$ between the system at $v = 1586~\kms$ towards 3C 273 and a dwarf galaxy at approximately the same $z$.  The dwarf galaxy has magnitude $M_B = -13.9$, and metallicity $Z = 0.10 \Zsun$, compared to $0.06 \Zsun$ for the absorber, derived along the absorber sightline \citep{Stocke04}.  The $z = 0.0777$ system towards PHL 1811 is consistent with a similar metallicity of $Z \sim 0.1 \Zsun$.  At $z = 0.0777$, and assuming $\Omega_\Lambda$ = 0.7, $\Omega_m$ = 0.3, and $H_0 = 70~\kms \textrm{Mpc}^{-1}$, the distance modulus would be $m - M = 37.7$.  Therefore, if the host  galaxy for PHL 1811 z = 0.0777 was similar to the dwarf galaxy described in \citet{Stocke04}, it would appear at magnitude $m_B = 23.9$.  It would thus be well below the magnitude limit of $m_R = 21.5$ in the \citet{Jenkins03} initial survey, and would not have been detected.

\subsection{PHL 1811 $z = 0.0777$ and other metal line absorbers}
\label{sec:OtherMetalLines}
 If the system at $z = 0.0777$ towards PHL 1811 is a condensed structure associated with a galaxy, it is likely to be related to other classes of absorbers in galactic environments.  A common tracer of denser gas in galactic haloes is \MgII.  If it is in a less dense phase, galactic halo gas appears as \CIV~absorption.  Finally, at very low densities, the gas shows up as \OVI~absorption lines.  Other lines may be used to discern the same phases; for example, \CII~and \SiII~for cooler gas, and \SiIV~for hotter gas.  The exact temperature and densities highlighted by each line depends on the photoionizing background, which in turn depends on the redshift.

Absorption in \CIII~would trace intermediate phase gas, with physical conditions between gas probed by \MgII~lines and gas probed by \CIV~lines.  However, the $z = 0.0777$ system towards PHL 1811 has two distinct components, with different inferred physical conditions.  Could either of these clouds be identified with known phases in galactic haloes?

\subsubsection{Analogs at $z \sim 0$: \MgII~and \CIV/\OVI}
The low redshift IGM is not as well studied as the high redshift IGM, because many of the important lines lie in the far ultraviolet.  Most low redshift studies concentrate on the overall statistics of absorbers, in particular, their number distribution and the fraction of the IGM they comprise.  None the less, some studies have considered which IGM phase is representative of each ion \citep[e.g.][]{Danforth06,Danforth08}.

It is clear that Cloud A is a member of a small subset of the \CIII~absorbers studied by \citet{Danforth06}.  This is evident simply based on the number of absorbers with detected weak {\SiII} and {\CII} absorption ($d{\cal N}/dz = 1.00\pm0.20$ for $0.02 < W_r(1260)<0.3$~{\AA}; \citealt{Narayanan05}) relative to {\CIII} ($d{\cal N}/dz = 10^{+3}_{-2}$ for $W_r(1335)>0.03$~{\AA}; \citealt{Danforth08}).  This subset is characterised by higher densities compared to the typical photoionized clouds described by \citet{Danforth06}, i.e. a lower ionization parameter.  The typical ionization parameter for photoionized \CIII~clouds is in the range $-3 < \log U < -1$ and we have found Cloud A to be at the very lower end of this range to explain the observed \CII~absorption.

Our favoured Cloudy photoionized cloud models predict a \MgII~rest frame equivalent width of $0.029$~{\AA} for Cloud A ($\log U = -3.0$) and a negligible value for the more highly ionized Cloud B ($\log U = -1.0$).  A value of $W_r(2796) = 0.029$~{\AA} would have been detected at high spectral resolution if the data had covered that spectral range.  As noted by \citet{Narayanan05}, only some of the E230M spectra in the STIS archive had high $S/N$ data for wavelengths 2800 -- 3110 \Ang~so that a total of $\sim 3$ weak \MgII~absorbers might be expected in the small total path length.  Because of this, direct knowledge of weak \MgII~absorbers at low redshift is limited.  In the case of this $z=0.0777$ absorber, the only spectrum covering the relevant near-UV region was obtained with the STIS G230L grating \citep{Jenkins03} and does not have a resolution sufficient for detection of a weak \MgII~line. However, the \SiII~1260 and \CII~1335 equivalent widths of the absorber imply that Cloud A is the equivalent of a weak \MgII~absorber \citep{Narayanan05}.  So, this subclass of \CIII~systems may largely coincide with the class of weak \MgII~systems at $z\sim 0$.

Although Cloud B does not have low ionization absorption, it is detected in \CIV~and can be compared to \CIV~absorbers.  \citet{Danforth08} studied very low redshift ($z \la 0.1$) \CIV~absorbers that were slightly stronger than Cloud B ($>30~\mAng$ compared to $26 \pm 6~\mAng$ for Cloud B).  They found that there was little correlation between $N(\CIII)$ and $N(\CIV)$.  They interpreted this to mean that \CIII~and \CIV~arise in different phases.  Furthermore, there was some correlation between $N(\CIV)$ and $N(\OVI)$, suggesting that $\CIV$ was collisionally ionized, or at least of the same phase as \OVI.  Interestingly, there does seem to be \OVI~absorption associated with Cloud B, with an equivalent width of $45.1 \pm 0.4~\mAng$ according to our Voigt profile fits (\S~\ref{sec:Spectrum}).  This detection was confirmed by \citet{Tripp08} in their study of low-$z$ \OVI~absorption.  Our more detailed modelling of the system suggests that it is photoionized, though not necessarily associated with an individual galaxy (see \S~\ref{sec:DiscussionColIon}).  While \OVI~absorbers may be collisionally ionized, some may be photoionized \citep[for example,][]{Savage02,Prochaska04}.  

We also note that \citet{Chen01} discovered $\sim 100~\kpc$ haloes in low redshift galaxies that appeared as \CIV~absorption.  In their picture, galaxies have haloes that become more rarefied and ionized with increasing radius.  So, the inner regions of galactic haloes would contain high density absorbers, such as those associated with \MgII, while the outer regions of would produce \CIV.  The inferred size of Cloud B ($83^{+65}_{-36}~\kpc$) in our favoured photoionized model, is of the same order as their extended haloes.  Since the density of Cloud A already suggests that it is associated with a galaxy, the \citet{Chen01} scenario would present a unified framework for interpreting the system: Cloud A is an overdensity in the inner halo of a galaxy, contained in Cloud B, the outer halo.  However, the \citet{Chen01} study was for $L_*$ galaxies, which should have been detected if present.  For this scenario to apply, the galaxy involved would have to be of lower luminosity.

In short, for this system, \CIII~coincides with both \MgII~and \CIV~phases.  However, neither Cloud A nor Cloud B seem to be completely typical \CIII~absorbers in the \citet{Danforth08} samples.  Instead, typical \CIII~absorbers would have conditions intermediate between Cloud A and Cloud B.  This implies that \CIII~actually traces several distinct phases of the IGM, and \CIII~absorbers are not a homogeneous population in terms of underlying physical conditions.  

\subsubsection{Analogs at $z \sim 1$: \CIV~and \OVI}
\label{sec:z1Analogs}
More information about galactic absorbers and their physical conditions is available at higher redshift, because the lines are redshifted into more convenient wavelengths.  Since the photoionizing background changes with redshift, becoming more intense at high $z$, we should not expect that the system would look the same at $z \sim 1$.  To compare the system at $z = 0.0777$ towards PHL 1811 with $z \sim 1$ systems, we used Cloudy to produce a simulated spectrum at $z = 1$ given the same physical conditions.  We used our favoured model ($\log U = -3$ and $\log Z = -1$ for Cloud A and $\log U = -1$ and $\log Z = -1$ for Cloud B) at $z = 0.0777$ as the baseline.  Density, size, metallicity, and abundance were kept the same.  However, the cloud was subject to the $z=1$ Haardt \& Madau spectrum, so that both the resulting ionization parameter and background ionizing spectrum were different.  The $z \sim 1$ model was in photoionization equilibrium, as was our favoured model.  

\begin{figure}
\includegraphics[width=84mm]{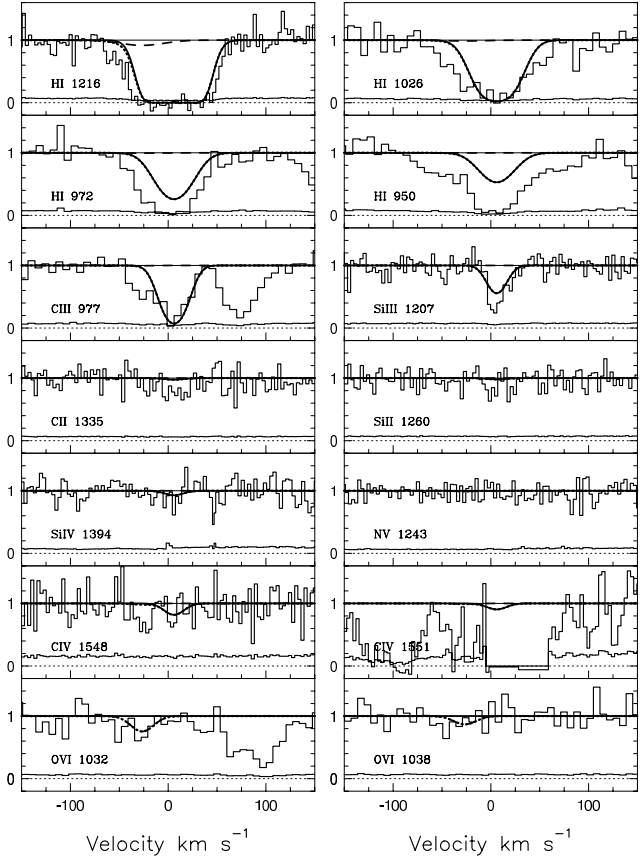}
\caption{Our favoured two cloud model of the $z = 0.0777$ absorber towards PHL 1811, as it would have appeared at $z = 1.0$.  As in Fig.~\ref{fig:ABAttr}, cloud A is solid and cloud B is long-dashed, with the total absorption as the dotted line.  For comparison, the actual low-$z$ spectrum is also shown.}
\label{fig:z1Spectrum}
\end{figure}

We show the simulated $z \sim 1$ spectrum in Fig.~\ref{fig:z1Spectrum}.  Cloud A, the denser component, appears as absorption in \HI, \CIII~and \CIV.  The \SiIII~absorption remains, but has grown weaker.  Similarly, the higher order Lyman series lines are also weaker.  There is also some very weak \SiIV~absorption.  However, the \CII~absorption has disappeared entirely.  At $z \sim 1$, then, Cloud A does not display low ionization lines but remains a \CIII~absorber.  Cloud B, the less dense component, is only weakly detected, and only in \Lya~and \OVI~absorption.  The \OVI~absorption has about the same strength as at $z = 0.0777$, but the \Lya~line has nearly vanished (compare with Fig.~\ref{fig:S4S6}), because more of the hydrogen has been ionized.   According to our simulated spectrum, \CIII~absorption would still come from the same phase as the denser \CIII~clouds at $z \sim 0$, those like cloud A.  However, at high redshift, higher ionization clouds like Cloud B would cease displaying \CIII.  Not only does \CIII~appear to come from several different phases at low redshift, but the set of phases it traces changes with redshift.  This will complicate studies of \CIII~that span a large range in $z$. 

Systems with similar spectral features are known to be associated with galaxies at $z \sim 1$.  Again, \CIV~and \MgII~are commonly used tracers of highly ionized and moderately ionized gas.  Cloud A, with a predicted radius of hundreds of parsecs, and a predicted density of $\sim 0.001~\cc$, is comparable to the inferred sizes and densities of $z \sim 1$ high ionization \CIV~clouds.  Furthermore, at $z \sim 1$, while high ionization clouds do occur without low ionization clouds at the same velocity, the converse is rarely true \citep{Milutinovic05}.  Clouds with the same physical conditions as these $z \sim 1$ are expected to manifest as \MgII, \CII, and \SiII~absorption at low redshift \citet{Narayanan05}.   Indeed, Cloud A does show \CII~absorption at low redshift, just as \CIV~absorbers at $z \sim 1$ are expected to.  

There are clear differences between this system and typical single cloud weak \MgII~absorbers at $z \sim 1$.  The $z \sim 1$ absorbers are often found, with photoionization modelling, to have smaller sizes and higher densities than any of the clouds in the $z = 0.0777$ system towards PHL 1811.   \citet*{Rigby02} constrained the iron-rich weak \MgII~absorbers to have low ionization parameters, high density ($\sim 0.1~\cc$), and thicknesses of only $\sim 10$ pc \citep{Rigby02}.  Those absorbers with little iron, like the $z = 0.0777$ absorber towards PHL 1811, were less well constrained, with sizes that could be larger than 10 pc and densities that could be lower than 0.1 \cc.  \citet{Charlton03} inferred depths of 0.1 to 100 pc, with average densities near 0.01 \cc.  In comparison, the predicted density of $\sim 0.001~\cc$ for our favoured model for Cloud A is quite small, and its size of $420^{+330}_{-180}$~pc is large.  Based on the inferred physical conditions of Cloud A and its predicted absorption at $z \sim 1$, we expect that Cloud A would have been a \CIV~absorber at those times, possibly of the same kind that envelops \MgII~absorbers at that redshift.  However, it would not have been a weak \MgII~absorber at $z \sim 1$.

\citet{Narayanan05}, however, note a problem with \CIV~phases at $z = 1$ becoming lower ionization phases at $z = 0$.  Because $W_r(2796)$ increases with time for a given system, more weak \MgII~absorbers would be expected to be detectable at present.  Their survey of \MgII~analogs, based on \CII~and \SiII~lines, suggests that $d{\cal N}/dz \sim 1$ for weak \MgII~absorbers at $z = 0$.  This implies that they are more rare in the present than in the past.  Since high ionization phases are expected to contribute more $z \sim 0$ \MgII~absorbers than low ionization phases, this suggests that the objects producing the \CIV~phases seen at $z = 1$ are evolving away by $z \sim 0$.  This could be because weak \MgII~absorbers may be persistent but shedding their high ionization material with time, or because the high ionization phases are transient and are being regenerated less often in the present.  However, our modelling of the system at $z = 0.0777$ towards PHL 1811 suggests that at least a few of these high ionization clouds survive to (or continue to be formed in) the present.

On the other hand, if cloud A corresponds to the high ionization phase found in weak \MgII~systems, then what is cloud B?  Cloud B appears to be still larger, less dense, and even higher ionization.  It seems to be another phase beyond those considered by \citet{Milutinovic05}, enshrouding the \CIV~components.  At high redshift, this phase may be visible in \OVI~or \Lya.  It may be some kind of photoionized WHIM in which cloud A and any host galaxy is embedded, as discussed in \S \ref{sec:DiscussionColIon}.  This scenario is consistent with photoionization modelling done by \citet{Zonak04} on the system at $z = 1.04$ on the PG 1634+706 line of sight.  That system is a `multicloud weak \MgII~absorber', with low ionization \MgII~clouds, intermediate ionization \SiIII~clouds, higher ionization \SiIV~clouds, and still higher ionization broad \OVI~components \citep{Zonak04}.  If the \SiIV~components at the same velocity as the \MgII~components are from a separate phase, then the \SiIV~clouds are best fit with sizes of several kiloparsecs, resembling Cloud A \citep{Zonak04}.  However, the broad \OVI~and \Lya~absorption in that system is fit by clouds with sizes of 12 \kpc~and 257 \kpc, comparable to our predictions for cloud B \citep{Zonak04}.   Our ($\log U = -1.0$, $\log Z = -1.0$) model produces the small detected \OVI~absorption feature from cloud B.   Compared with cloud B, the \OVI~absorption is stronger in \citet{Zonak04} with $W_r$ being a factor of $\sim 2 - 3$ times greater for each of their \OVI~clouds, because those clouds had greater total column density, as well as higher metallicity or ionization parameter \citep{Zonak04}.   

\subsubsection{The environments of \CIII~and other metal line absorbers}
\label{sec:Environment}

\citet{Rosenberg03} studied a pair of sightlines that pass through the Virgo Cluster, 3C 273 and RX J1230.8+0115.  They found several \Lya~absorbers at the redshift of the Virgo cluster, with ionization parameters of $\log U \sim -2.8$, conditions similar to those for weak \MgII~absorbers, and detected weak \CII~and \SiII~ absorption \citep{Rosenberg03}.  Photoionization modelling implied sizes of a few tens of kiloparsecs \citep{Rosenberg03}.  None the less, absorbers were found to be `correlated' between the sightlines, though the separation was nearly $350 h^{-1}~\kpc$ \citep{Rosenberg03}.  \citet{Rosenberg03} were sceptical that the absorbers were part of `contiguous' filaments, because of velocity dispersions within the Virgo cluster and because the absorbers on the two sightlines were not identical.  They also found that only one of the absorbers, the v = 1589 \kms~system towards 3C 273, could be associated with a galaxy \citep{Rosenberg03}.  So, they suggested a scenario in which the absorbers were part of different overdensities in the same large scale structure.  The overdensities could be from dwarf winds, though this explanation requires a large amount of initial \HI~gas in the dwarfs \citep{Rosenberg03}.  This appears to be a similar situation to the $z = 0.0777$ system on the PHL 1811 sightline, which is embedded in known large scale structure filaments, near several galaxy clusters, but has no identifiable host.

When considering single cloud weak \MgII~absorption systems at $z < 1$, \citet{Milutinovic05} use a statistical argument to show that the high ionization \CIV~phases are identified with the $100 h^{-1}(L/L_B)^{0.5}~\kpc$ \CIV~($W_r(1548) \ge 0.1~\Ang)$ haloes seen around bright galaxies \citep{Chen01}.  Our favoured model of the system at $z = 0.0777$ on the PHL 1811 sightline argues against this conjecture, at least as applied to $z = 1$ \CIV~absorbers.  Cloud A seems to be a descendant of the high ionization phase surrounding weak \MgII~absorbers seen at $z \sim 1$, corresponding to \CIV~clouds with sizes of order a half a \kpc.  On the other hand, cloud B is probably a \CIV~cloud (at $z \sim 0$) with a size of tens of \kpc, similar to the size of the \CIV~haloes inferred by \citet{Chen01}.  Cloud B, therefore, itself seems to be reminiscent of the \CIV~haloes described in \citet{Chen01} -- yet, at $z \sim 1$ with more intense ionizing radiation it would be an \OVI~absorber.  So, the higher ionization \CIV~phases seen in weak \MgII~absorbers at high redshift, which are similar to Cloud A, may not be the \CIV~haloes at low redshift, which are similar to Cloud B.  Instead, the \CIV~absorbers seen at $z \sim 1$ (similar to Cloud A) may be overdensities in relatively spherical haloes (including Cloud B).  Alternatively, the large \CIV~haloes may be large scale structure associated with the host galaxies of more compact absorbers, like cloud A, or denser, filamentary \CIV~absorbers at $z \sim 1$.

\section{Conclusion}
\label{sec:Conclusion}
We have modelled the $z = 0.0777$ \CIII~absorption system towards PHL~1811 in order to derive its physical conditions and to compare to other classes of absorbers.  We find the system to be best fit by two photoionized \CIII~clouds, separated by $\sim 35$~\kms, with $\log Z \sim -1$: Cloud A, with $\log U \sim -3$, which also gives rise to \SiIII~and weak \CII~absorption, and Cloud B, with $\log U \sim -1$, which also gives rise to weak \CIV~and \OVI~absorption (see Fig.~\ref{fig:S4S6}). These parameters are constrained to within $\sim \pm 0.25$~dex. Photoionization models explain these two clouds better than collisional ionization models, both because they can simultaneously produce all observed metal line transitions and because collisional ionization models have very short cooling times.

For our favoured model, cloud A has a density of $n_e = 1.2^{+0.9}_{-0.5} \times 10^{-3}$~{\cc} and a thickness of $0.4^{+0.3}_{-0.2}$~kpc.  If we had coverage of the appropriate wavelength range, we would expect to observe weak \MgII~absorption from this cloud, with $W_r(2796) \sim 0.03\ \Ang$.  We predict with a photoionization model (in Fig.~\ref{fig:z1Spectrum}) that, had it existed at $z \sim 1$, Cloud A would have appeared as \CIII, \SiIII, and \CIV~absorption, while cloud B would have been an \OVI~absorber.  Despite its identity as a weak, low ionization absorber at low redshift, Cloud A has a lower density and large size than the large population of weak low ionization absorbers studied at $z\sim 1$ through \MgII~absorption.  The Cloud A properties are similar to those of \CIV~absorbers (without detected low ionization) at $z\sim 1$, but the decreased extragalactic background radiation at low redshift leads to a shift towards a lower ionization state.  Cloud B has a density of $n_e = 1.2^{+0.9}_{-0.5} \times 10^{-5}$~{\cc} and a thickness of $80^{+70}_{-40}$~kpc, suggesting that it may be analogous to \OVI~absorbers at $z \sim 1$.

Physically, we suggest that the two \CIII~clouds~in this absorber are significantly condensed regions, with cloud A (overdensity $\delta \sim 5000$ to within a factor of $2$) in some way related to a galaxy environment, and with cloud B (overdensity $\delta \sim 50$ to within a factor of $2$) embedded in the large scale structure.  The two clouds are different phases of the IGM.  Neither seems typical of \CIII~absorbers: Cloud A shows the rarer \CII, while Cloud B shows \CIV~and \OVI, which \citet{Danforth08} argues is from a different IGM phase than \CIII.  We conclude that \CIII~absorbers in general represent several IGM phases, including the two phases of our system.

\section*{Acknowledgments}
This research was funded by NASA under grant NAG5-6399 NNG04GE73G.  We are grateful to R. Lynch and J. Masiero for their work with the STIS archive spectra.  We also would like to thank the people, especially C. Danforth, who made useful suggestions at the 207th Meeting of the American Astronomical Society in January 2006.

\end{document}